%% file: main.tex
\documentclass[prd,preprint,nofootinbib]{revtex4-1}
\pdfoutput=1
\usepackage[latin1]{inputenc}
\usepackage[T1]{fontenc}
\usepackage{amsmath}
\usepackage{float}
\usepackage{graphicx}
\usepackage{comment}
\usepackage{caption}
\usepackage{subcaption}
\usepackage{xcolor}
\usepackage{color}
\usepackage{epsfig}
\usepackage{dcolumn}
\usepackage{subfig}
\usepackage{caption}
\usepackage{subcaption}
\usepackage{bm}
\usepackage{amssymb}
\usepackage{slashed}

\begin{document}

\title{Modification of the B Meson Mass in a Magnetic Field from QCD Sum Rules}

\author{C.~S.~Machado}
\email{camilasm@ift.unesp.br}
\affiliation{Instituto de F\'isica Te\'{o}rica, Universidade Estadual Paulista, SP, Brazil}

\author{S.~I.~Finazzo}
\email{stefanofinazzo@gmail.com}
\affiliation{Instituto de F\'{i}sica, Universidade de S\~{a}o Paulo, S\~{a}o Paulo, SP, Brazil}

\author{R.~D.~Matheus}
\email{matheus@ift.unesp.br}
\affiliation{Instituto de F\'isica Te\'{o}rica, Universidade Estadual Paulista, SP, Brazil}

\author{J.~Noronha}
\email{noronha@if.usp.br}
\affiliation{Instituto de F\'{i}sica, Universidade de S\~{a}o Paulo, S\~{a}o Paulo, SP, Brazil}

\date{\today}

\begin{abstract}
In this paper we extend the well known QCD sum rules used in the calculation of the mass of heavy mesons to estimate the modification of the charged B meson mass, $m_B$, in the presence of an external Abelian magnetic field, $eB$. Two simplifying limits were considered: the weak field limit in which the external field satisfies $e B \ll m^2$ (with $m$ being any of the masses involved) and the strong field limit in which the field strength is small in comparison to the bottom quark mass (or the B meson mass) squared but it is large compared to the mass of the light quarks, i.e., $m^2_{u,d} \ll e B \ll m^2_{b,B}$. We found that $m_B$ decreases with the magnetic field in the both of these limits.

\end{abstract}

\maketitle

\input{introduction}

\input{qcdsumrules}
\input{qcdsumrulesmag}
\input{numericalres}

\input{conclusions}
\input{appendix}

\input{bibliography}
\end{document}

%% file: introduction.tex
\section{Introduction}

Strong magnetic fields can be relevant to a number of physical systems. Some dense stars, such as highly magnetized neutron stars known as magnetars, can display magnetic fields as large as $eB \sim 1~\mathrm{MeV}^2$~\cite{Duncan:1992hi}\footnote{The field strength in the CGS system is $(eB)^2\sim 59.14 \times 10^{-22}\,(B/1\,{\rm G})$ GeV$^2$, i.e, a magnetic field of magnitude 1.69 $\times \,10^{20}$ G corresponds to 1 GeV$^2$.}. Of a more direct interest to particle physics are the electromagnetic fields produced in ultrarelativistic heavy ion collisions performed at the Large Hadron Collider (LHC) or at the Relativistic Heavy Ion Collider (RHIC). Non-central collisions in these colliders can produce short-lived electromagnetic fields where the intensity of the magnetic field can be as large as $eB \sim m_{\pi}^2 \sim 0.02~\mathrm{GeV}^2$ (at RHIC) or $eB \sim 15\, m_{\pi}^2 \sim 0.3~\mathrm{GeV}^2$ (at LHC)~\cite{m1,m2}, where $m_{\pi}$ is the pion mass.

These field strengths are comparable with the hadronic mass scale and could have important phenomenological implications to the physics of hadrons in Quantum Chromodynamics (QCD). Much effort has been given towards understanding the effects of strong magnetic fields on the different phases of the QCD diagram \cite{book} and, recently, lattice QCD simulations with physical quark masses have determined how the deconfinement and chiral phase transitions are affected by strong magnetic fields \cite{Bali:2011qj}.

The mass spectrum of the hadronic states, however, is set by the non-perturbative regime of QCD and one has to look for the appropriate non-perturbative tools to describe it and include the effects of the magnetic field therein. For instance, one would expect that magnetic fields with values defined at the hadronic scale would modify the binding energy of the various hadronic states, which could then affect their masses.

Various different methods have been successful in dealing with the hadronic spectrum over the years such as, for instance, quark potential models \cite{Lichtenberg:1987ms}, lattice QCD \cite{Davies:2003ik,Fodor:2012gf}, and the QCD Spectral Sum Rules (QCDSR)~\cite{QCDSR, SNB}. Weak external fields in QCDSR were introduced in the past in order to analyze the magnetic moments of hadrons \cite{Ioffe:1983ju,Ioffe:1983cj}. In this work we propose a novel way to include an external magnetic field into the QCDSR formalism for the two-point correlators in order to study its effect on the masses of scalar $B^{\pm}$ mesons. We start in Section \ref{qcdsumrules.sec} with a short review on the QCDSR in the vacuum (i.e., in the absence of external fields, $eB = 0$). The external field contributions are considered in Section \ref{QCDSRMag.sec} and they enter the QCDSR in two ways.

First, the effects of an external magnetic field on the quark propagators are taken into account using the non-perturbative Schwinger propagator \cite{schwinger}. This modification describes how the perturbative sector of QCD changes due to the magnetic field. The complete calculation using the proper-time propagator is technically difficult and we discuss a series of approximations that allow us to extract results in some limiting situations. The second modification introduced in this paper with respect to the usual QCDSR approach is the dependence of the condensates that parametrize the non-perturbative character of QCD with the magnetic field. The effect of magnetic fields on the chiral condensate was evaluated by a number of means, including chiral perturbation theory \cite{condcampo1,condcampo2,condcampo3}, Nambu-Jona-Lasinio models \cite{condcampo4, shovkovy}, and lattice QCD \cite{Bali:2012zg, rede1, rede2, rede3, rede4, rede5}. Using condensates that are functions of the field strength we take into account the effects of the magnetic field on the long distance, non-perturbative interactions of QCD.

The numerical analysis of the QCDSR and the results for the masses are shown in Section IV, where we show that the magnetic field has the effect of lowering the masses of the heavy mesons studied, which is in agreement with recent potential model calculations \cite{simonov,machado}.

%% file: qcdsumrules.tex
\section{QCD Sum Rules in the vacuum}

\label{qcdsumrules.sec}

QCDSR \cite{QCDSR, SNB} are based on the evaluation of the two point correlation function in the vacuum
\begin{equation}
\label{eq:correlator}
\Pi(q) = i \int d^4 x \,e^{iq\cdot x} \langle 0|T\{j(x)j^{\dagger}(0)\}| 0 \rangle ,
\end{equation}
where $j(x)$ is an interpolating current carrying the quantum numbers of the hadron in question. For the $B^{\pm}$ mesons, we will use
\begin{equation}
\label{eq:current}
j(x)=\bar{q}_a(x) i \gamma_5 Q_a(x),
\end{equation}
where $q$ is the light quark field, $Q$ is the heavy quark field, and $a$ is a color index. With this current in \eqref{eq:correlator}, the correlator can be written as
\begin{align}
\label{eq:correlatorpseudo}
\Pi(q)= \frac{i}{(2\pi)^4}\int d^4k\, \text{Tr}[S_{ab}^q(k) \gamma_5 S_{ba}^Q(k+q) \gamma_5].
\end{align}
where $S^q$ ($S^Q$) is the full propagator for a light (heavy) quark.

Based on the principle of quark-hadron duality, which states that correlation functions of colorless currents in QCD can be described either in terms of quarks and gluons or hadronic degrees of freedom, the correlator (\ref{eq:correlator}) will be evaluated here in two different ways. On the one hand, we start from a perturbative description based on the quark and gluon degrees of freedom using Wilson's Operator Product Expansion (OPE) \cite{Wilson:1969zs} to evaluate (\ref{eq:correlator}) in the presence of nonzero vacuum condensates that act as a source for non-perturbative effects. The resulting expression for the correlation function is called the OPE side. On the other hand, on the phenomenological side, we use a description based on hadronic degrees of freedom by inserting a complete set of hadronic states in (\ref{eq:correlator}) to obtain the correlator in terms of a dispersion relation
\begin{align}
\label{eq:dens}
\Pi^{\text{phen}}(k)=\int_0^{\infty} ds \frac{1}{s-k^2-i\epsilon} \rho(s),
\end{align}
where $\rho(s)$ is the spectral density. The following parametrization is generally used for the spectral density
\begin{align}
\label{eq:polecont}
\rho^{\text{phen}}(s)=
\frac{m_H^4}{m_Q^2} f_H^2
\delta(s-m^2_H)+\theta(s-s_0)\rho^{\text{cont}}(s),
\end{align}
where $m_H$ is the mass of the ground state of the hadron, $s_0$ is the continuum threshold and $f_H$ is the coupling of this state with the current, which for heavy-light mesons is defined by $\langle 0 | j |H \rangle = \frac{m_H^2}{m_Q + m_q} f_H \approx \frac{m_H^2}{m_Q} f_H  $, and $m_Q$ and $m_q$ are the masses of the heavy and light quarks, respectively. This parametrization separates the contribution of the lowest lying pole from that of the excited states, collectively called ``the continuum''. The parameter $s_0$ indicates when the excited states start to contribute significantly to the spectral density.

We can also write the OPE side in terms of a dispersion relation $\rho^{\text{OPE}} \equiv \frac{\mathrm{Im}\, \Pi^{\text{OPE}}}{\pi}$ and the principle of quark-hadron duality allows us to assume
$\rho^{\text{cont}} = \rho^{\text{OPE}}$. Then, the phenomenological side takes the form
\begin{align}
\label{eq:fen}
\Pi^{\text{phen}}(k) = \frac{m_H^4 f_H^2}{m_Q^2 \left(m_H^2-k^2\right)}+ \int^{\infty}_{s_0} ds \frac{\rho^{\text{OPE}}(s)}{s-k^2},
\end{align}
while the OPE side can be written as
\begin{align}
\label{eq:OPE}
\Pi^{\text{OPE}}(k)=\int^{\infty}_{s_{\text{min}}} ds \frac{\rho^{\text{OPE}}(s)}{s-k^2},
\end{align}
where $s_{min} = (m_q+m_Q)^2$. Taking the Borel transform and imposing the quark-hadron duality at the level of correlators $\hat{\Pi}^{OPE} (\bar{M}^2) = \hat{\Pi}^{phen} (\bar{M}^2)$, where $\bar{M}$ is the Borel mass~\cite{QCDSR, SNB}, we arrive at the sum rule
\begin{align}
\label{eq:vacQCDsumrule}
\frac{m_H^4}{m_Q^2} f_H^2
e^{-m_H^2/\bar{M}^2}= \int_{s_{\text{min}}}^{s_0} ds \, \rho^{\text{OPE}}(s)\, e^{-s /M^2}.
\end{align}
Taking the derivative of Eq.\ (\ref{eq:vacQCDsumrule}) with respect to $1/\bar{M}^2$ and dividing the resulting expression by (\ref{eq:vacQCDsumrule}) we get an explicit expression for the hadron mass
\begin{align}
\label{eq:vacmass}
m^2_H = \frac{\int_{s_{\text{min}}}^{s_0} ds \, \rho^{\text{OPE}}(s) \, s \, e^{-s/\bar{M}^2}}{\int_{s_{\text{min}}}^{s_0} ds \, \rho^{\text{OPE}}(s) \, e^{-s/\bar{M}^2}}.
\end{align}

In our calculations we will consider the OPE expansion up to operators of dimension 3. It is known~\cite{Reinders:1981ty, Shuryak:1981fza, Narison:1987qc} that for heavy states such as the B meson the contribution of higher dimension condensates is small and, thus, such terms can be omitted in a first approximation. In the vacuum ($eB = 0$) this means that one needs to consider only the identity operator and the quark condensate $\langle \bar{q} q \rangle$ in the calculations.

%% file: qcdsumrulesmag.tex
\section{QCD Sum Rules With Magnetic Fields}

\label{QCDSRMag.sec}

In order to determine the contributions from the OPE to the correlation function in the presence of an external magnetic field we will consider, as the ``free" quark propagator, the full non-perturbative propagator computed by Schwinger \cite{schwinger} that describes the interaction of a spin 1/2 field with the magnetic field. However, this propagator does not include the QCD interactions experienced by the quarks, which are parametrized here using the non-perturbative QCD condensates. The interactions of the magnetic field with the QCD vacuum are taken into account by considering the dependence of the condensates with the external field. On the phenomenological side, we will take the pole contribution as being given by the full propagator of a charged scalar meson in an external magnetic field.

\subsection{Quark propagator in the presence of an external magnetic field}

The Schwinger proper-time representation \cite{schwinger} describes the Feynman propagator of a spin $1/2$ fermion with charge $e$ and mass $m$ in an external, constant and uniform Abelian magnetic field. Considering the magnetic field in $z$ direction and the symmetric gauge, i.e $\mathbf{A}=(-B y/2,B x/2,0)$, the Schwinger propagator can be written as \footnote{We use a mostly minus signature for the Minkowski metric and  4-vectors $v^\mu\equiv(v_0,v_1,v_2,v_3)$ are separated into parallel, $v_{\parallel} \equiv (v_0,0,0,v_3)$, and perpendicular pieces, $v_{\perp} \equiv (0,v_1,v_2,0)$, with respect to the direction of the magnetic field. The inner product is written as $u_\mu v^\mu \equiv u\cdot v = u_{\parallel} \cdot v_{\parallel} - u_{\perp} \cdot v_{\perp}$, where $u_{\parallel} \cdot v_{\parallel} \equiv u_0 v_0 - u_3 v_3$ and $u_{\perp} \cdot v_{\perp} \equiv u_1 v_1 + u_2 v_2$. Thus, for instance, $u_{\parallel}^2 = u_0^2 - u_3^2$ and $u_{\perp}^2 = u_1^2 + u_2^2$.}
\begin{align}
\label{eq:propschwinger}
S_{ab}(k)= \delta_{ab}\int_0^{\infty} d s \exp\left[i s\left(k_0^2 - k_3^2 - k_{\perp}^2 \frac{\tan(eB s)}{eB s} - m^2\right) \right] \times \nonumber &\\
\times  \left[(k^0\gamma^0-k^3\gamma^3+m)(1+\gamma^1\gamma^2\tan(eBs))
-  k_{\perp}\cdot \gamma_{\perp}(1+\tan^2(eBs)) \right].&
\end{align}

We can also write the propagator as a sum over Landau levels~\cite{shovkovy}
\begin{equation}
\label{eq:propsumoverlandau}
S_{ab}(k) = i \delta_{ab} e^{-k_{\perp}^2/|eB|} \sum\limits_{n=0}^{\infty} (-1)^n \frac{D_n(eB,k)}{k_{\parallel}^2-m^2-2|eB|n+i\epsilon},
\end{equation}
with
\begin{align}
\label{eq:propsumoverlandau2}
D_n(eB,k)  = (k^0\gamma^0-k^3\gamma^3+m) \left[(1-\gamma^1 \gamma^2 \mathrm{sign}\,(eB) ) L_n \left( \frac{2 k_{\perp}^2}{|eB|} \right) + \right. & \nonumber\\
\left. - (1+i\gamma_1 \gamma_2 \mathrm{sign}\,(eB)) L_{n-1} \left( \frac{2 k_{\perp}^2}{|eB|} \right) \right] + 4 (k^1\gamma^1+k^2\gamma^2) L^1_{n-1} \left( \frac{2 k_{\perp}^2}{|eB|} \right)&
\end{align}
where $L^a_n$ are the associated Laguerre polynomials and $L_n \equiv L^0_n$ \footnote{For $n<0$ one defines $L_n = L^1_{n-1} = 0$.}. The form \eqref{eq:propschwinger} of the propagator is more convenient when considering weak fields ($eB \ll m^2$), as it can be easily expanded in powers of $eB/m^2$. The alternative form \eqref{eq:propsumoverlandau2} is convenient when one is interested in a strong field limit ($eB \gg m^2$), since in this case the Lowest Landau Level (LLL), given by $n=0$, dominates. We shall develop later in this section how these approximations are relevant to the study of the heavy mesons considered in this paper.

\subsection{Non-perturbative QCD contributions}

The Schwinger propagator in Eq.\ (\ref{eq:propschwinger}) takes into account in a non-perturbative manner all the effects coming from the external magnetic field on the quark propagators but we have not taken into account the intrinsic non-perturbative QCD effects (and their modification due to the external field).
In the QCDSR method, the non-perturbative aspects of QCD are accounted for by performing the OPE of the correlator and considering the vacuum expectation values of the local operators thus obtained.

Schematically, one can think of an expansion on the quark propagators themselves\footnote{Note that the expansion is performed within the correlator and one must be careful with OPE terms that potentially involve more than one propagator, such as the gluon condensate. However, since we are only considering condensates of dimension 3, such subtleties do not appear in our calculations.} and write
\begin{equation}
\label{eq:quarkprop1}
S^q_{ab,\alpha \beta} = S^{q,\text{pert}}_{ab,\alpha \beta} + \langle : q_{a \alpha} (x) \bar{q}_{b \beta} (0) : \rangle,
\end{equation}
where $S^{q,\text{pert}}_{ab,\alpha \beta}$ is the perturbative propagator, which in the presence of a magnetic field corresponds to \eqref{eq:propschwinger}. As for the normal ordered term, up to condensates of dimension 3, we get (see Appendix \ref{app.ope.sec} for further details)
\begin{equation}
\label{eq:quarknormal}
\langle : q_{a \alpha} (x) \bar{q}_{b \beta} (0) : \rangle = -\frac{\delta_{ab}}{12} \langle : \bar{q} q : \rangle \delta_{\alpha \beta} - \frac{\delta_{ab}}{12} \langle : \bar{q} \sigma_{1 2} q : \rangle \sigma^{1 2}_{\alpha \beta}.
\end{equation}

The non-perturbative QCD effects are parametrized by the condensates $\langle \bar{q} q \rangle$ and $\langle \bar{q} \sigma_{1 2} q \rangle$. The inclusion of the effects from the magnetic field on the condensate terms will be done by taking $\langle \bar{q} q \rangle$ and $\langle \bar{q} \sigma_{1 2} q \rangle$ as functions of $eB$. The value used for the light quark chiral condensate in the absence of the magnetic field is $\langle \bar{q} q \rangle_0= (-0.23)^3 \, \mathrm{GeV}^3$ while $\langle \bar{Q} Q \rangle \sim 0$ for heavy quarks \cite{x3872,narpdg}. For the ratio $\Sigma(eB) \equiv \langle \bar{q} q \rangle (eB)/\langle \bar{q} q \rangle_0$ we used two different parametrizations according to the intensity of the magnetic field. For magnetic fields such that $eB/m_\pi^2 \ll 1$ we use the chiral perturbation theory result \cite{condcampo1,condcampo3}
\begin{equation}
\label{eq:weakchiralcond}
\Sigma(eB) = 1 + \ln 2\,\frac{ eB}{16 \pi^2 F_{\pi}^2} I_H \left(\frac{m_{\pi}^2}{eB} \right),
\end{equation}
where $F_{\pi} \sim 93 \, \mathrm{MeV}$ is the pion decay constant, $m_{\pi} = 140 \, \mathrm{MeV}$ is taken to be the pion mass, and
\begin{equation}
\label{eq:integraldef}
I_H(y) = \frac{1}{\ln 2} \left[ \ln (2 \pi) + y \ln \left( \frac{y}{2}\right) - y - 2 \ln \Gamma \left(\frac{1+y}{2} \right) \right].
\end{equation}
For magnetic fields $eB > 1$ GeV$^2$, we used a linear extrapolation of the lattice results in \cite{Bali:2012zg} (note, however, that the contribution from the condensates to the mass of $B$ mesons is very small and, thus, our final results for the masses are not sensitive to such an extrapolation). In the limit of weak fields, the lattice results are compatible with those found in chiral perturbation theory \cite{Bali:2012zg}. Due to the Dirac matrix structure of the pseudo-scalar current in Eq.\ \eqref{eq:current}, the term proportional to $\langle \bar{q} \sigma_{1 2} q \rangle$ results in a vanishing trace in \eqref{eq:correlatorpseudo} and, thus, such term does not enter in our calculations. However, for other types of interpolating currents (such as the vector mesons $\Upsilon(1S)$ or $B^{*}$) the contribution from the $\langle \bar{q} \sigma_{1 2} q \rangle$ condensate may enter explicitly in the OPE and must be considered. With this possibility in mind, we remark that there are already lattice results for this quantity in the presence of a magnetic field \cite{Bali:2012jv}.

The non-perturbative contribution to the correlator that comes from using the propagator \eqref{eq:quarkprop1} in the OPE two point function, after the Borel transform, is
\begin{align}
\hat{\Pi}_{\mathrm{\langle \bar{q} q \rangle}} (\bar{M}) = -m_Q \langle \bar{q} q \rangle e^{-m_Q^2/\bar{M}^2},
\label{borelqq}
\end{align}
where $m_Q$ is the heavy quark mass and $\bar{M}$ is the Borel mass.

\subsection{Weak field approximation}

In this paper the weak field limit is defined by the condition $eB \ll m_q^2 \ll m_{Q,B}^2$, where $m_q$ is the mass of the light quark and $c$ is the charge of the light quark in units of the electron charge. This considerably simplifies the propagator in Eq.\ (\ref{eq:propschwinger}). We define $x \equiv s m^2$ (where $m$ here can be either one of the masses) and expand \eqref{eq:propschwinger} in powers of $(ceB/m^2)$, up to quadratic order, obtaining ($eB > 0$)
\begin{align}
S_{ab}(k) = \delta_{ab} \int_0^{\infty} dx \frac{ e^{i x \alpha/m^2}}{m^2}\left[(\slashed k + m)  + (k_{\parallel} \cdot \gamma_{\parallel}+m)\gamma^1\gamma^2 \left(\frac{ceB x}{m^2} \right) - k_{\perp} \cdot \gamma_{\perp}\left(\frac{ceB x}{m^2} \right)^2\right] ,&
\end{align}
with
\begin{align}
\alpha \equiv k^2-m^2-\frac{k_{\perp}^2}{3}\left(\frac{ceB x}{m^2}\right)^2.
\end{align}
Evaluating the integrals, we obtain
\begin{align}
\label{eq:propweak}
S(k)= S^{(eB)^0}(k) + S^{(eB)^1}(k) + S^{(eB)^2}(k),
\end{align}
with
\begin{align}
S^{(eB)^0}(k) = i\frac{-\slashed k +m}{k^2-m^2}
\end{align}
and
\begin{align}
S^{(eB)^1}(k) = \left(\frac{eB}{m^2}\right)\left[-(k_{\parallel} \cdot \gamma_{\parallel}+m)\gamma^1\gamma^2\frac{m^2}{(k^2-m^2)^2} \right] ,
\end{align}
\begin{align}
S^{(eB)^2}(k) = \left(\frac{eB}{m^2}\right)^2\left[-2ik_{\perp}^2\frac{m^4}{(k^2 - m^2)^4} + 2ik_{\perp}\cdot\gamma_{\perp}\frac{m^4}{(k^2-m^2)^3}\right].
\end{align}

For the perturbative part of the
quark propagators we use (\ref{eq:propweak}), which is a good approximation as long as $eB \ll m_q^2 \ll m_Q^2$. These propagators are then inserted in the correlation function \eqref{eq:correlatorpseudo}. The momentum integrals are evaluated using a Feynman parametrization and a cutoff regularization - the details can be found in Appendix B. In the end, we arrive at

\begin{align}
\Pi (q) = \Pi^{(eB)^0}_{\text{pert}} + \Pi_{1,\text{pert}}^{(eB)^2} + \Pi_{2,\text{pert}}^{(eB)^2} + \Pi_{\langle \bar{q} q \rangle},
\end{align}
where
\begin{align}
\Pi^{(eB)^0}_{\text{pert}} = \frac{3 \cdot }{(2\pi)^2} \int_0^1 \mathrm{d}x \left[ 2 \Delta - x(1-x) q^2 -m_q m_Q \right] \log \Delta,
\end{align}
\begin{align}
\label{pi0}
\Pi_{1,\text{pert}}^{(eB)^2} = \frac{3}{(2\pi)^2} (cC)(eB)^2\int_0^1 \mathrm{d}x (1-x)x \left[ \frac{1}{\Delta} - \frac{(x^2-x)q_\parallel^2+m_q m_Q}{2\Delta^2}\right],
\end{align}
\begin{align}
\Pi_{2,\text{pert}}^{(eB)^2} = \frac{3\cdot 2}{(2\pi)^2} (c^2+C^2)(eB)^2\int_0^1 \mathrm{d}x \left\{\frac{x^3}{3} \left[ \frac{3}{2\Delta} + \frac{m_q m_Q + q_{\perp}^2(3-7x+4x^2)}{2\Delta^2} + \right. \right.  &\nonumber \\  \left. \left. -\frac{(x^2-x)q^2}{2\Delta^2}+(1-x)^2q_{\perp}^2 \frac{m_q m_Q -(x^2-x) q^2}{\Delta^3} + x^2 \left[\frac{1}{\Delta}+ \frac{(x^2-x)q_{\perp}^2}{\Delta^2} \right]  \right] \right\}&,
\end{align}
where $C$ is the charge of the heavy quark in units of the electric charge and
\begin{align}
\Delta \equiv (x^2-x)q^2+x m_q^2 + (1-x)m_Q^2.
\end{align}
One can check that all linear terms in $eB$ have vanishing trace.

One may think that the expressions above could be easily simplified by taking the limit in which $m_q \rightarrow 0$. However, this approximation is not strictly allowed in the weak field limit since in this case $eB \ll m_q^2 \ll m_Q^2$. However, it is possible to rewrite the equations in terms of the dimensionless parameters  $m_q/m_Q$ and $eB/m_Q^2$ to show that the terms proportional to $m_q$ only contribute to the real part of the correlator. Since we are only interested in $\rho^{\mathrm{OPE}}(q)$, which comes from the imaginary part of the correlator, we can safely take $m_q \rightarrow 0$ in this case. The result is an integral with a logarithmic term whose branch cut yields the imaginary part of the correlator plus polynomial terms without an imaginary part.

When $m_q \rightarrow 0$ the kinematic constraint in the $s=q^2$ integral is $s_{\text{min}} = q_{\text{min}}^2 =  m_Q^2$. We can use the relation $q^2 = q_{\parallel}^2 - q_{\perp}^2$ to choose two of the three momenta as independent variables - it will be useful to choose $s=q^2$ and $q_{\perp}^2$. In the end, the spectral density of the OPE side is given by
\begin{equation}
\rho_{\mathrm{pert}}^{\text{weak}}(s=q^2,q_{\perp}^2) = \rho^{(eB)^0}(s,q_{\perp}^2) + \rho^{(eB)^2}_1(s,q_{\perp}^2) + \rho^{(eB)^2}_2(s,q_{\perp}^2) 
\label{eq:rhoweak}
\end{equation}
with
\begin{equation}
\rho^{(eB)^0}(s,q_{\perp}^2) = \frac{3}{8\pi^2}\left[\frac{(s-m_q^2)^2}{s} + 2\frac{m_Q m_q}{s} \left(s -m_Q^2\right)\right],
\end{equation}
\begin{equation}
\rho^{(eB)^2}_1(s,q_{\perp}^2) =  - \frac{3}{4\pi^2}(cC)(eB)^2\frac{m_Q^2}{s^3} \left(q_{\perp}^2\right),
\end{equation}
and
\begin{equation}
\rho^{(eB)^2}_2(s,q_{\perp}^2) = - \frac{(c^2+C^2)(eB)^2 m_Q^4}{4\pi^2(m_Q^2-s)^2(s)^3}\left[6m_Q^4 -m_Q^2(s-14 q_{\perp}^2)-3s(2s+7q_{\perp}^2) \right]\,.
\end{equation}
The Borel transformed correlator in the weak field approximation is
\begin{equation}
\hat{\Pi}_{\text{weak}}^{\mathrm{OPE}}(\bar{M},q_{\perp}^2) = \int_{s_{\text{min}}}^{s_0} ds\;\left[ \rho_{\mathrm{pert}}^{\text{weak}}(s,q_{\perp}^2) \, e^{-s/\bar{M}^2} \right] + \hat{\Pi}_{\mathrm{\langle \bar{q} q \rangle}} (\bar{M}). 
\label{eq:piweak}
\end{equation}

\subsection{Strong field approximation}

In the opposite limit, i.e., very strong magnetic fields such that $m_q^2 \ll eB \ll m_Q^2$, we can use the alternative representation (\ref{eq:propsumoverlandau}) for the light quark propagator and keep only the lowest Landau level when $ceB/m_q^2 \gg 1$ \cite{shovkovy}. In other words, we truncate the Landau sum for the light quark propagator at $n=0$
\begin{align}
\label{eq:propLLL}
S^{(0)}_{ab}(k)=i \delta_{ab} \,e^{-k_{\perp}^2/(ceB)} \left(\frac{k_{\parallel} \cdot \gamma_{\parallel}+m_q}{k_{\parallel}^2-m_q^2}\right)\left(1-i\gamma^1\gamma^2 \right).
\end{align}
With respect to the heavy quark mass the magnetic field is not strong, $eB \ll m_Q^2$, and we can still use the Taylor expansion in Eq.\ (\ref{eq:propweak}).

With these propagators, we obtain 
\begin{align}
\Pi(q) = \Pi_{\text{pert}}^{\text{strong}}+ \Pi_{1,\text{pert}}^{\text{strong}}+\Pi_{2,\text{pert}}^{\text{strong}}+\Pi_{\langle \bar{q} q \rangle},
\end{align}
where
\begin{align}
\Pi_{\text{pert}}^{\text{strong}}(q)
= 3 \cdot 4 i \int \frac{d^2k_{\perp}}{(2\pi)^2} e^{-\frac{k_{\perp}^2}{ceB}}
\int \frac{d^2k_{\parallel}}{(2\pi)^2} \frac{k_{\parallel}^2+ k_{\parallel} \cdot q_{\parallel}}{((k+q)^2-m_Q^2)k_{\parallel}^2}
\end{align}
and
\begin{align}
\Pi_{1,\text{pert}}^{\text{strong}}(q)
= -3 \cdot 4 i \frac{eB}{m_Q^2}\int \frac{d^2k_{\perp}}{(2\pi)^2} e^{-\frac{k_{\perp}^2}{ceB}}
\int \frac{d^2k_{\parallel}}{(2\pi)^2} \frac{k_{\parallel}^2+ k_{\parallel} \cdot q_{\parallel}}{((k+q)^2-m_Q^2)^2k_{\parallel}^2}.
\end{align}
One can show that the term proportional to $(eB/m_Q^2)^2$, $\Pi_{2,\text{pert}}^{\text{strong}}(q)$, vanishes after taking the Dirac trace. The $k_{\parallel}$ integral can be done with the Feynman parametrization. The integral in the Feynman parameter results in a logarithmic term only for $\Pi_{\text{pert}}^{\text{strong}}(q)$, from which we extract the imaginary part and, thus, the OPE spectral density. The $\Pi_{1,\text{pert}}^{\text{strong}}(q)$ integral is real and does not contribute to the imaginary part. Therefore, the spectral density of the OPE side in this strong field limit is given by
\begin{align}
\rho_{\text{pert}}^{\text{strong}}(s=q^2,q_{\perp})= \frac{3}{2\pi}e^{-q_{\perp}^2/(ceB)} \int_0^{\sqrt{s+q^2_{\perp} - m_Q^2}} dk_{\parallel}k_{\parallel}e^{-\frac{k_{\perp}^2}{ceB}}I_0\left(\frac{2k_{\parallel}\sqrt{q_{\perp}^2+s}}{ceB} \right).
\label{eq:opestrong}
\end{align}

Finally, the Borel transformed correlator in the strong field approximation is
\begin{align}
\hat{\Pi}_{\text{strong}}^{\text{OPE}}(\bar{M},q_{\perp}) = \int_{s_{\text{min}}}^{s_0} ds\;\left[ \rho_{\mathrm{pert}}^{\text{strong}}(s,q_{\perp}) \, e^{-s/\bar{M}^2} \right] + \hat{\Pi}_{\mathrm{\langle \bar{q} q \rangle}} (\bar{M}). 
\label{eq:pistrong}
\end{align}

\subsection{The phenomenological side}

Since we are dealing with a charged (pseudo)scalar meson, we can use the Schwinger propagator for a spin 0 particle to describe the pole that appears in the phenomenological part of the QCDSR
\begin{align}
\label{propscalar}
G(q) = -i \int_0^{\infty} \frac{ds}{\cos(eBs)} \exp \left[-is\left(m_H^2 - q_{\parallel}^2 + \frac{\tan(eBs)}{eBs} q_{\perp}^2 \right) \right],
\end{align}
here $m_H$ is the mass of the hadronic state (in the present case, $m_H=m_B$).

Although this propagator is fully non-perturbative with respect to the external magnetic field, its full form is rather complicated to implement in the evaluation of the correlation function (\ref{eq:correlator}). Since $m^2_{H} \gg eB$ in both scenarios explored on the OPE side, in this paper we expand the charged pseudoscalar propagator in powers of $eB/m_H^2$ in the phenomenological part of the QCDSR. Given that magnetic fields of the order $eB \sim m^2_{\pi}$ \cite{m1,m2,toneev,Tuchin:2013ie} are the most relevant for the study of heavy ion collisions at RHIC and LHC, in \cite{wip} the calculations presented in this paper will be generalized to consider effects from magnetic fields of arbitrary strength.

Therefore, we expand the propagator in \eqref{propscalar} in powers of $eB/m_H^2$ and the final result up to order $(eB)^2$ is
\begin{align}
\label{eq:schwingerscalar}
G(q) = \frac{1}{q^2-m_H^2} - (eB)^2 \left[\frac{1}{(q^2-m_H^2)^3} + \frac{2 q_{\perp}^ 2}{(q^2-m_H^2)^4} \right].
\end{align}

The Borel transform of the pole phenomenological side is then given by
\begin{align}
\hat{\Pi}^{\text{phen}}(\bar{M}^2) = \frac{m_H^4}{m_Q^2} f_H^2 e^{-m^2_H/\bar{M}}\left[1 - (eB)^2\left(\frac{1}{2\bar{M}^4} + \frac{q_{\perp}^2}{3\bar{M}^6} \right) \right]\,.
\label{eq:borelphen}
\end{align}

%% file: numericalres.tex
\section{Numerical Results}

\subsection{Vacuum}\label{vacuumsection}

Since we are interested in the effect of the magnetic field on the meson mass, we will normalize our calculations by its vacuum ($eB=0$) value also determined via QCDSR. Thus, we will briefly review the numerical results for the mass of the charged B meson computed via QCDSR in the absence of a magnetic field, which was already studied in~\cite{Reinders:1981ty, Shuryak:1981fza, Narison:1987qc}.

We fixed $\langle q\bar{q} \rangle = (-0.23)^3~\mathrm{\text{GeV}}$ and the quark masses $m_b = 4.24~\mathrm{\text{GeV}}$ and $m_{u,d} \approx 0$. These values were chosen to mantain consistency with other QCDSR calculations~\cite{x3872,narpdg}. The continuum threshold $s_0$ is a free parameter fixed using the phenomenological rule $(m_H + 400) \, \mathrm{\text{MeV}} \lesssim \sqrt{s_0} \lesssim (m_H + 800) \, \mathrm{\text{MeV}}$.

\begin{figure}
\begin{center}
\begin{subfigure}{.5\textwidth}
    \includegraphics[width=.95\linewidth]{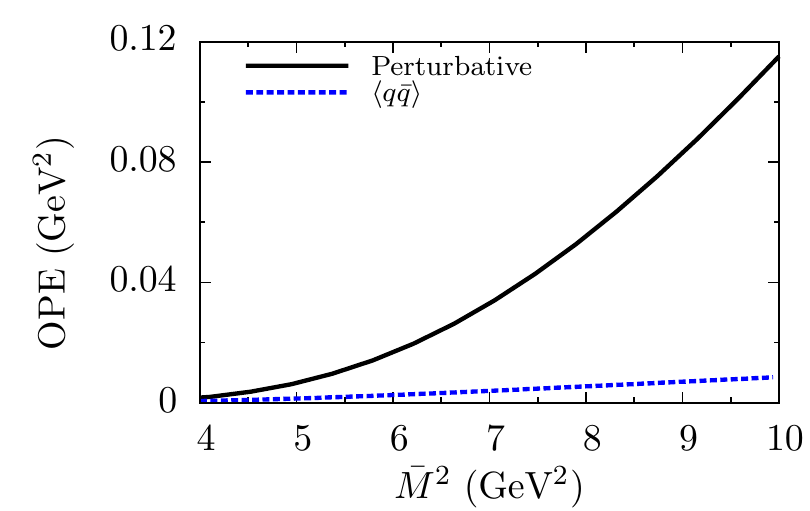}
    \caption{}
  \label{fig:vacuo}
\end{subfigure}%
\begin{subfigure}{.5\textwidth}
   \includegraphics[width=.95\linewidth]{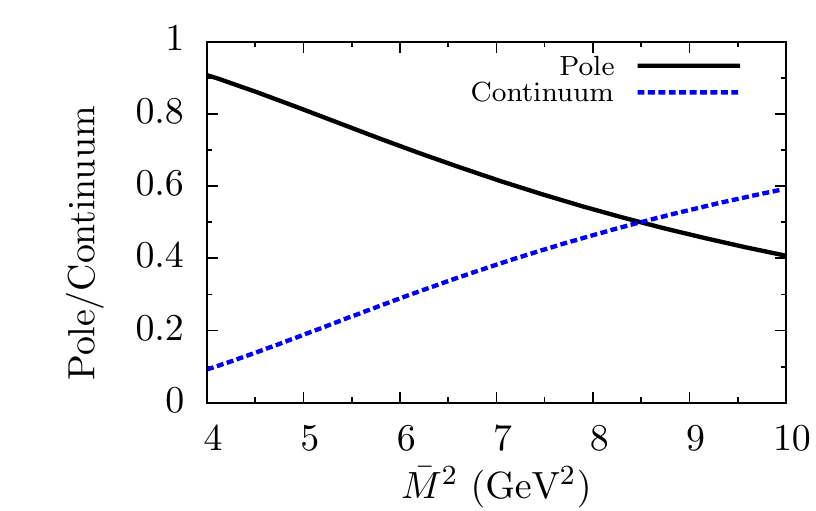}
   \caption{}
  \label{fig:vacuo2}
\end{subfigure}
\end{center}
\caption{(a) Convergence of the OPE expansion and (b) pole dominance in the absence of a magnetic field.}
\label{fig:test}
\end{figure}

In the QCDSR approach, there is an interplay between the convergence of the OPE (valid for large squared momentum $q^2$ or low Borel mass $\bar{M}^2$) and the contribution from the continuum of excited states (which become very important for low $q^2$ or large $\bar{M}^2$). The OPE convergence is estimated by requiring that the contribution from the condensates of dimension 3 is less than $10 \%$ of the perturbative contribution - this gives a lower limit to the Borel mass $\bar{M}_{\text{min}}$ (see Fig.\ \ref{fig:vacuo}). An upper limit $\bar{M}_{\text{max}}$ is determined by requiring that the contribution from the pole is larger than that from the continuum 
(see Fig.\ \ref{fig:vacuo2}). We obtain  $\bar{M}_{\text{min}}^2 \sim 4~\mathrm{\text{GeV}}^2$ (Fig.\ \ref{fig:vacuo}) and $\bar{M}_{\text{max}}^2 \sim 8~\mathrm{\text{GeV}}^2$ for $\sqrt{s_0}=6.0~\mathrm{\text{GeV}}$.
The interval determined by $\bar{M}_{\text{min}}$ and $\bar{M}_{\text{max}}$ is called the Borel window and the procedure explained above is used to fix it throughout this work.

\begin{figure}
\begin{center}
\begin{subfigure}{.5\textwidth}
    \includegraphics[width=.95\linewidth]{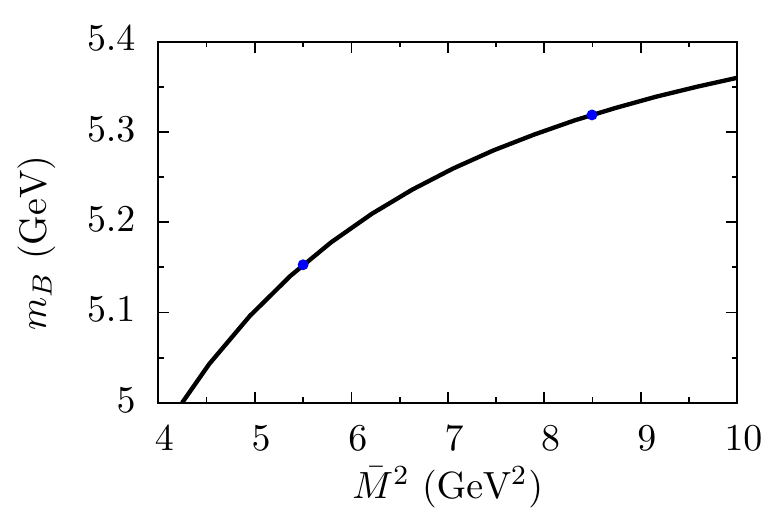}
    \caption{}
  \label{fig:vacuo3}
\end{subfigure}%
\begin{subfigure}{.5\textwidth}
   \includegraphics[width=.95\linewidth]{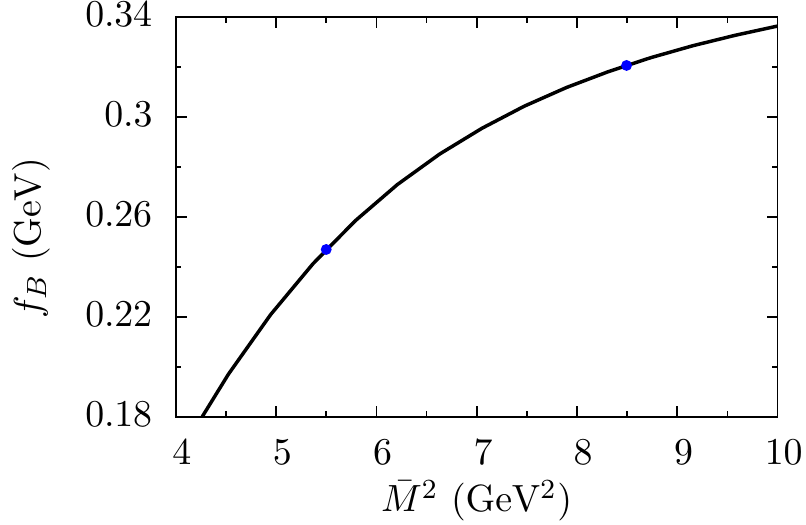}
   \caption{}
  \label{fig:vacuo4}
\end{subfigure}
\end{center}
\caption{Dependence of the mass (a) and coupling constant (b) of the B meson, as functions of the Borel mass, in the absence of a magnetic field. The points indicate the Borel window.}
\label{fig:test2}
\end{figure}

With the Borel window fixed, we can determine the hadronic parameters (mass and coupling of the B meson) by averaging the values in the Borel window (Fig.\ \ref{fig:vacuo3} for the mass and \ref{fig:vacuo4} for the coupling). We obtain $m_B = 5.25~\mathrm{\text{GeV}}$ and $f_B = 0.29~\mathrm{\text{GeV}}$, which agree with the experimental values~\cite{Beringer:1900zz} 
and the results in~\cite{Reinders:1981ty, Shuryak:1981fza}.

The errors in the QCDSR calculations come mainly from the truncation of the OPE expansion, from the choice of the continuum threshold ($s_0$), and from the uncertainties in the values of the quark masses and condensates. These errors can be estimated by varying those parameters within their uncertainties. In this work, we will not make an estimate of QCDSR intrinsic errors since we are interested in the ratios between sum rules calculations with and without the effects of magnetic fields, where these errors are expected to cancel out.

\subsection{Weak magnetic fields}

The sum rules for the weak magnetic field case are obtained by equating Eqs.~(\ref{eq:piweak}) and (\ref{eq:borelphen})
\begin{align}
\label{eq:weakqcdsr}
\hat{\Pi}_{\mathrm{\langle \bar{q} q \rangle}} (\bar{M}) + \int_{s_{min}}^{s_0} ds e^{-s/\bar{M}^2} \rho_{\mathrm{pert}}^{\text{weak}}(s,q_{\perp})=
\frac{m_H^4}{m_Q^2} f_H^2
e^{-m^2_H/\bar{M}^2}\left[1 - (eB)^2\left(\frac{1}{2\bar{M}^4} + \frac{q_{\perp}^2}{3\bar{M}^6} \right) \right]
\end{align}
where $\hat{\Pi}_{\mathrm{\langle \bar{q} q \rangle}} (\bar{M})$ is given by \eqref{borelqq}. In Eq.\ \eqref{eq:weakqcdsr}, the parameters we want to discover (for a given Borel mass $\bar{M}$) are the meson mass $m_H$ and the coupling constant $f_H$. As in the vacuum case, by differentiating \eqref{eq:weakqcdsr} with respect to $1/\bar{M}^2$ we can obtain a second equation to solve for $m_H$ and $f_H$. However, due to the more complex dependence of the phenomenological side on $\bar{M}$, we cannot eliminate $f_H$ from Eq.\ \eqref{eq:weakqcdsr} by the same procedure done in the vacuum. Thus, we numerically solve \eqref{eq:weakqcdsr} and its derivative with respect to $1/\bar{M}^2$ to obtain simultaneously $m_H$ and $f_H$. As a consistency check, we verified that this procedure yields the same numerical results found for the vacuum in Section \ref{vacuumsection}.

\begin{figure}
\begin{center}
\begin{subfigure}{.5\textwidth}
    \includegraphics[width=.95\linewidth]{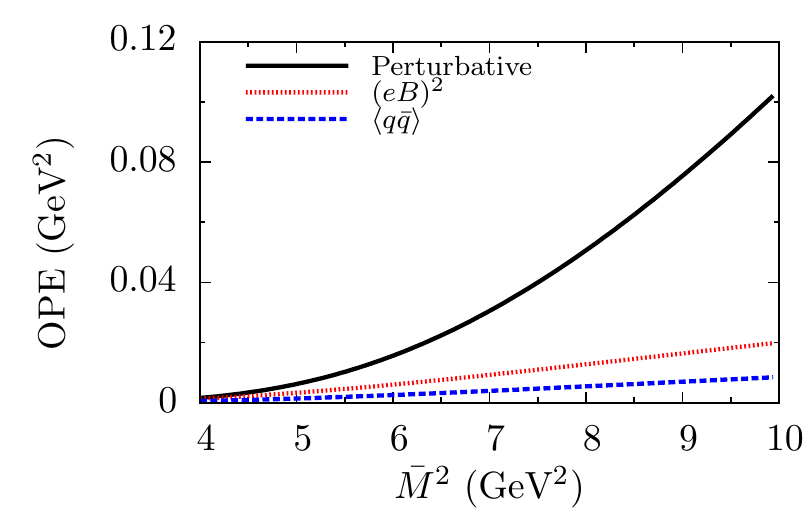}
    \caption{}
  \label{fig:weakfield}
\end{subfigure}%
\begin{subfigure}{.5\textwidth}
   \includegraphics[width=.95\linewidth]{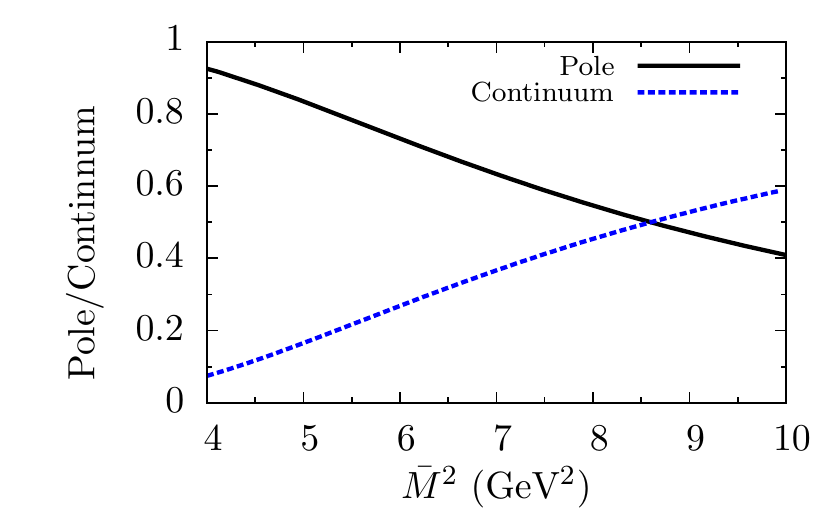}
   \caption{}
  \label{fig:weakfield2}
\end{subfigure}
\end{center}
\caption{(a) OPE convergence and (b) pole dominance for $eB = 4 \cdot 10^{-8}~\text{GeV}^2$ and $q_{\perp}^2 =1~\text{GeV}^2$.}
\label{fig:test3}
\end{figure}

\begin{figure}
\begin{center}
    \includegraphics[width=.475\linewidth]{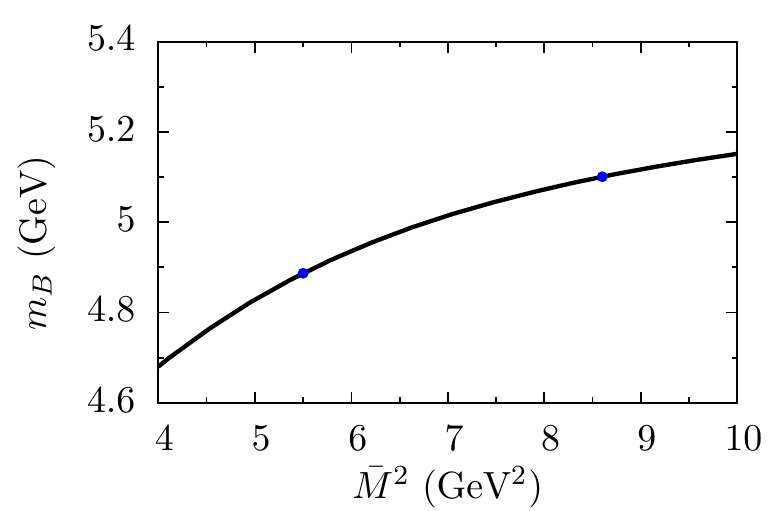}
\end{center}
\caption{Mass of the B meson, $m_B$, as a function of the Borel mass $\bar{M^2}$ for $eB = 4 \cdot 10^{-8}~\text{GeV}^2$ (weak field limit) and $q_{\perp}^2 =1~\text{GeV}^2$. The points indicate the Borel window.}
\label{fig:weakfield3}
\end{figure}

For this initial study, we fix  $eB = 2 \cdot 10^{-6} m_{\pi}^2 \sim 4 \cdot 10^{-8}~\text{GeV}^2$ and $q_{\perp}^2 =1~\text{GeV}^2$ - this last choice reflects a typical hadronic scale. 
The convergence of the OPE in Fig.\ \ref{fig:weakfield} and the pole dominance can be seen in Fig.\ \ref{fig:weakfield2}. The result for $m_B$
for these fixed values of $eB$ and $q_{\perp}^2$ is shown in Fig.\ \ref{fig:weakfield3},
along with the respective Borel window. For magnetic fields larger than $ \sim 4 \cdot 10^{-8}~\text{GeV}^2$ the contribution from the term $\sim (eB)^2$ is larger than the vacuum term, signaling the breakdown of our weak field expansion for the light quark propagator.

\begin{figure}
\begin{center}
\begin{subfigure}{.5\textwidth}
    \includegraphics[width=.95\linewidth]{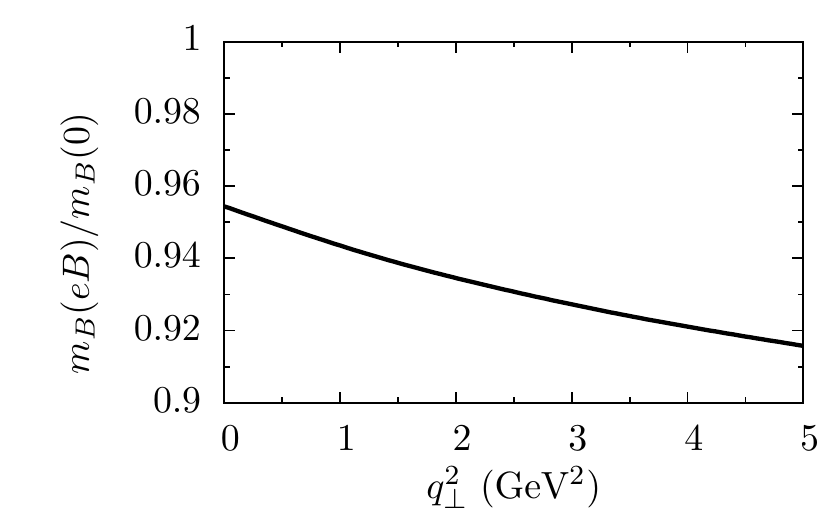}
    \caption{}
  \label{fig:weakfield5}
\end{subfigure}%
\begin{subfigure}{.5\textwidth}
   \includegraphics[width=.95\linewidth]{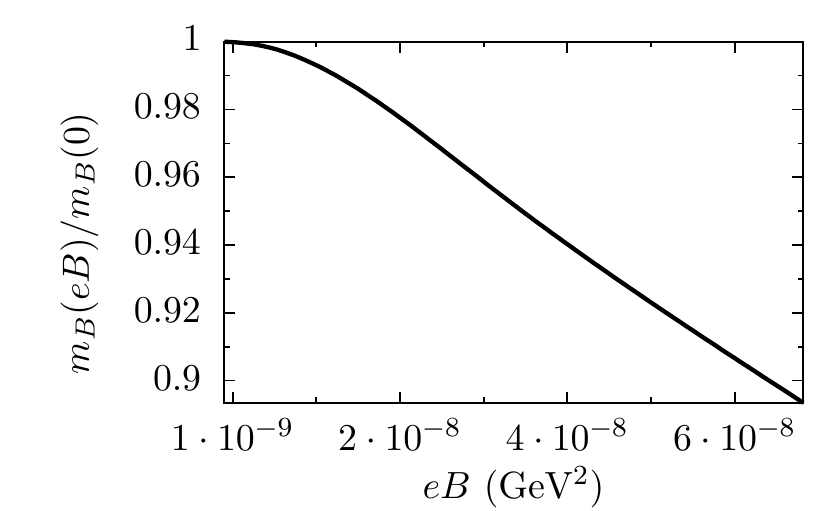}
   \caption{}
  \label{fig:weakfield6}
\end{subfigure}
\end{center}
\caption{(a) B meson mass, $m_B$, as a function of the perpendicular momentum for a fixed magnetic field of $eB=2\cdot 10^{-6} m_{\pi}^2$ (weak field limit) and (b) dependence of $m_B$ on the magnetic field for a fixed perpendicular momentum $q_{\perp}^2=1~\text{GeV}^2$.}
\label{fig:test5}
\end{figure}

A more systematic study can be done to investigate the role of the choice of $q_{\perp}^2$ by fixing $eB \sim 2\cdot 10^{-6} m_{\pi}^2$ 
and varying $q_{\perp}^2$. The results are shown in Fig.\ \ref{fig:weakfield5}, where the computed masses and couplings are normalized by the vacuum results.
One can see that $m_B$ is quite sensitive to the choice of $q_{\perp}^2$. Nevertheless, for any choice of $q_{\perp}^2$ the effect of the magnetic field is to lower $m_B$. This is consistent with the ``Zeeman" splitting found for the $\rho$ meson mass in the presence of magnetic fields \cite{simonov,hidaka,luschevskaya}. However, in the QCDSR approach we capture only the hadron ground state and, thus, one should expect to obtain only the lower meson mass.

With these observations in mind, we fixed $q_{\perp}^2$ to be $1~\text{GeV}^2$ and varied $eB$. The results are shown in Fig.\ \ref{fig:weakfield6}. One can see that $m_B$ decreases with increasing $eB$, as expected.

\subsection{Strong magnetic fields}

\begin{figure}
\begin{center}
\begin{subfigure}{.5\textwidth}
    \includegraphics[width=.95\linewidth]{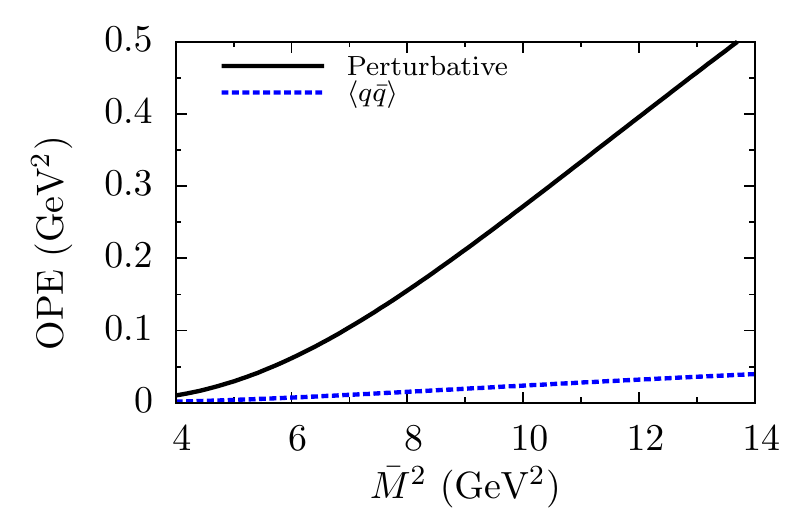}
    \caption{}
  \label{fig:strongfield}
\end{subfigure}%
\begin{subfigure}{.5\textwidth}
   \includegraphics[width=.95\linewidth]{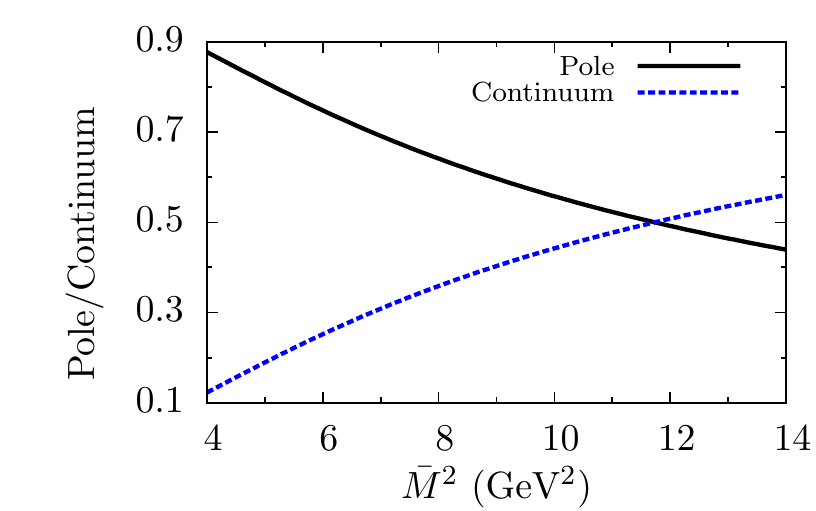}
   \caption{}
  \label{fig:strongfield2}
\end{subfigure}
\end{center}
\caption{(a) OPE convergence and (b) pole dominance for $eB = 1.5~\text{GeV}^2$ (strong field limit) and $q_{\perp}^2 =0.5~\text{GeV}^2$.}
\label{fig:test6}
\end{figure}

In the strong magnetic field limit, the calculation is entirely analogous to the one realized in the preceding subsection, except that now we use (\ref{eq:pistrong}) for the OPE side. In this case, we are in the limit $m_q^2 \ll eB \ll m_Q^2$. By varying $eB$, we see that to have a valid Borel window we have to limit $eB$ to be in the range $eB \sim 50 m_{\pi}^2 \sim 1~\text{GeV}^2$ and $eB \sim 200m_{\pi}^2 \sim 4~\text{GeV}^2$.
In Figs.\ \ref{fig:strongfield} and \ref{fig:strongfield2} we show the convergence of the OPE and the dominance of the pole over the continuum for $eB = 75  m_{\pi}^2 \sim 1.5~\text{GeV}^2$ and $q_{\perp}^2 =0.5~\text{GeV}^2$.

The continuum threshold, $s_0$, is one of the main sources of error in the QCDSR approach and, thus, one needs to be careful with the choice of this parameter in the strong field limit. Using the standard phenomenological estimate $(m_H + 400) \, \mathrm{\text{MeV}} \lesssim \sqrt{s_0} \lesssim (m_H + 800) \, \mathrm{\text{MeV}}$ as a guide, we chose, for $eB = 1.0 \, \mathrm{GeV}^2$, $eB = 2.5 \, \mathrm{GeV}^2$, and $eB = 4.0 \, \mathrm{GeV}^2$, three values of $s_0$ which satisfy $\sqrt{s_0} \sim (m_H + 600) \, \mathrm{\text{MeV}}$, using an interpolation of $s_0$ for intermediate values of $eB$. To analyze the sensitivity of the results with $s_0$, we also repeated this analysis for values of $\sqrt{s_0}$ in the range defined by $\sim (m_H + 400) \, \mathrm{\text{MeV}}$ and $\sim (m_H + 800) \, \mathrm{\text{MeV}}$, yielding a lower and upper limit curves of continuum thresholds, $\sqrt{s_0} (eB)$, $\sqrt{s_{0,min}}$ and $\sqrt{s_{0,max}}$, respectively.

\begin{figure}
\begin{center}
\begin{subfigure}{.5\textwidth}
   \includegraphics[width=.95\linewidth]{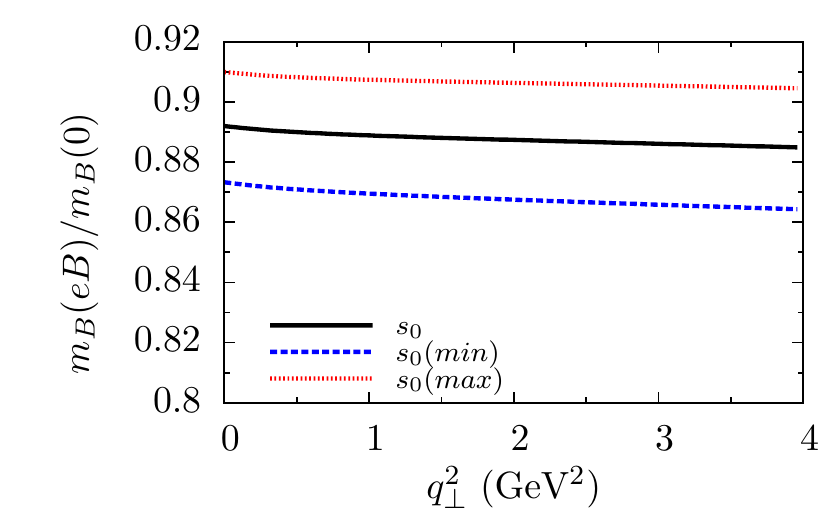}
   \caption{}
  \label{fig:strongfield10}
\end{subfigure}%
\begin{subfigure}{.5\textwidth}
    \includegraphics[width=.95\linewidth]{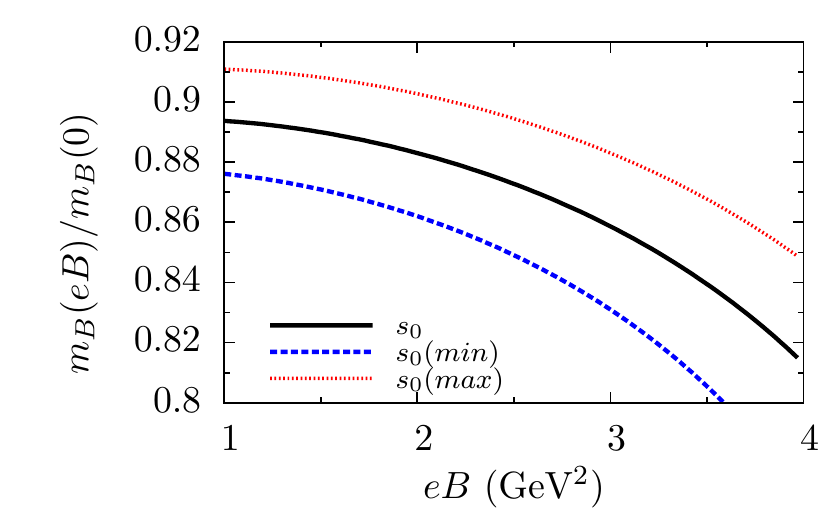}
    \caption{}
  \label{fig:strongfield9}
\end{subfigure}
\end{center}
\caption{(a) Mass of the B meson, $m_B$, as a function of the perpendicular momentum $q_{\perp}^2$ for fixed $eB = 1.5~\text{GeV}^2$ (strong field limit) and (b) as a function of the magnetic field $eB$ for fixed $q_{\perp}^2=0.5~\text{GeV}^2$. The curves correspond to the central $\sqrt{s_0} \sim (m_H + 600) \, \mathrm{\text{MeV}}$, lower $\sqrt{s_0} \sim (m_H + 400) \, \mathrm{\text{MeV}}$, and upper $\sqrt{s_0} \sim (m_H + 800) \, \mathrm{\text{MeV}}$ interpolated continuum thresholds $s_0 (eB)$ curves, as described in main text.}
\label{fig:test10}
\end{figure}

The results for the mass as a function of $eB$ and $q^2_{\perp}$ with the three values of $s_0$ are shown in Figs.\ (\ref{fig:strongfield10}) and (\ref{fig:strongfield9}).
Note that in the strong field limit $m_B$ displays the same qualitative behavior as a function of $eB$ as observed in the weak field case. However, in contrast with the weak field result (Fig.\ \ref{fig:weakfield6}), $m_B$ is found to be less sensitive to the choice of $q_{\perp}^2$.

%% file: conclusions.tex
\section{Conclusions}

In this paper we have introduced a modification of the QCDSR method in order to estimate the effects of external magnetic fields on the mass of charged B mesons. The effect of such fields has been taken into account via two distinct modifications. First, the quark propagators (on the OPE side) and the meson propagator (on the phenomenological side) were modified using the proper-time representation introduced by Schwinger~\cite{schwinger}, which gives the exact propagators for fermions and scalar particles in the presence of a constant and uniform magnetic field. Secondly, the quark condensate, which encodes the non-perturbative aspect of QCD, has been replaced by its magnetic field dependent value and the same approach could be used for higher dimensional condensates. While these modifications include all the effects of a constant magnetic field in the QCDSR method, the full calculation using the complete proper-time propagators is technically difficult to implement. In this first study, we decided to restrict the calculation to some limiting situations in order to gain qualitative insight and order of magnitude estimates of the possible effects. Two such simplifying limits were considered: (1) the ``weak field'' limit, where the external field satisfies $e B \ll m^2$ (with $m$ being any of the masses involved, both of the quarks and the meson) and (2) the ``strong field'' limit, where the field strength is still small compared to the bottom quark mass or the B meson mass squared, but still large enough compared to the light quarks, i.e., $m^2_{u,d} \ll e B \ll M^2_{b,B}$.

In the ``weak field'' limit we can expand all proper-time propagators in powers of $ceB/m^2$ (with $m$ being the mass of a given propagator), which greatly simplifies the calculation. We kept terms up to $(eB/m^2)^2$ and evaluated the QCDSR with condensates up to dimension $3$. Surprisingly enough, we have found sizable effects already with considerably weak fields ($eB \sim 2\cdot 10^{-6} m^{2}_{\pi}$). The effect of the field is to lower the meson mass. This result agrees with the expectation that the magnetic field splits the meson into two states, and the fact that the QCDSR only considers the lowest lying state. The surprising feature found here is the magnitude of the mass suppression, which is about $10\%$. We also found a strong dependence of the mass with respect to the meson momentum perpendicular to the magnetic field, which might have some phenomenological implications. Our calculations behave properly in the limit of $eB\rightarrow 0, q_{\perp} \rightarrow 0$, falling back to the usual QCDSR results.

In the ``strong field'' limit we considered fields of the order $eB \sim 75 m_{\pi}^2$. In this limit we can still expand the propagators of the heavy quark and the meson in the same way we did for the previous case. The light quark propagator, on the other hand, can be written as a sum over Landau levels and for such a strong field we assumed that only the lowest Landau level contributed significantly, which allowed us to truncate the sum to its first term. In this approximation we found that the decrease in the B meson mass is in the $10\%$ to $20\%$ range depending on the field strength, perpendicular momentum, and the intrinsic QCDSR parameters. This is not a large effect, especially considering the results found in the weak field limit. It seems that most of the magnetic field effects take place at smaller field values, with the mass changing at a slower pace after that. That might indicate the presence of a saturation mechanism that stabilizes the meson mass as a function of the magnetic field, but a more complete calculation (valid for arbitrary values of the magnetic field) is needed to verify if that is indeed the case.  

The values of magnetic fields relevant to ultrarelativistic heavy ion collisions, $eB \sim 1-15 \,m_{\pi}^2$, are in between the two sets of values considered in this paper. Although we do not expect the effects of the magnetic field to change qualitatively the result found here, a more complete calculation valid for arbitrary values of the magnetic field is needed to confirm this expectation. Such a calculation is also desirable since it could be used to study other mesons that do not have the same separation of scales present in B mesons (which justified our approximations). We are currently tackling the more general calculations that include the explicit sum over Landau levels and we intend to present the results in a future publication~\cite{wip}.

\begin{acknowledgments}
This work has been partly supported by Funda\c c\~ao de Amparo \`a Pesquisa do Estado de S\~ao Paulo (FAPESP) and Conselho Nacional de Desenvolvimento Cient\'ifico e Tecnol\'ogico (CNPq). C.~S.~M.\ is supported by FAPESP under contract 2011/05619-3 and 2012/21627-9, S.~I.~F.\ is supported by FAPESP under contract 2011/21691-6. The authors thank F.~S.~Navarra and I.~A.~Shovkovy for discussions about the QCD condensates in a magnetic field and M.~Nielsen and R.~Higa for comments on the manuscript.
\end{acknowledgments}

%% file: appendix.tex
\appendix

\section{The OPE for the quark propagator}

\label{app.ope.sec}

In this Appendix we work out the OPE expansion for the quark propagator in a constant, homogeneous magnetic field including the effects of condensates of dimension 3. The procedure we follow can be extended to include condensates of higher dimension. However, note that in general new condensates appear due to the magnetic field and for dimensions higher than 4 one does not yet have estimates for these new condensates in the presence of a magnetic field.

We start with the propagator for a quark field $q$
\begin{align}
\label{eq:quarkprop}
S^q_{ab,\alpha \beta} \equiv & \langle \Omega | T \left\{ q_{a \alpha} (x) \bar{q}_{b \beta}(0) \right\} | \Omega \rangle = \nonumber \\ = & \langle0 | T \left\{ q_{a \alpha} (x) \bar{q}_{b \beta}(0) \right\} | 0 \rangle + \langle \Omega | : q_{a \alpha} (x) \bar{q}_{b \beta} (0) : |\Omega \rangle ,
\end{align}
where $|\Omega \rangle$ is the true vacuum, $|0 \rangle$ is the perturbative vacuum, $T$ is the time ordering operator, $a,b=1,2,3$ are color indices and $\alpha=1,2,3,4$ is a Dirac matrix index. The first term in the second line is the perturbative propagator. In our case, instead of the free propagator we will use the Schwinger proper-time propagator for a fermion in a magnetic field, since it includes all the contributions from the external field (but is free from the point of view of QCD interactions). The second term, i.e., the normal ordered product, will be expanded in terms of the QCD condensates.

Our approach is analogous to the one used to obtain the quark propagator in the QCDSR method in nuclear matter (see, for example, \cite{Cohen:1994wm} for a review). The main idea is that one can expand the matrix element $\langle : q_{a \alpha} (x) \bar{q}_{b \beta} (0) : \rangle \equiv \langle \Omega | : q_{a \alpha} (x) \bar{q}_{b \beta} (0) : |\Omega \rangle$ in terms of the usual basis for the Dirac matrices, $\{1, \gamma_{\mu}, \gamma_5, \gamma_5 \gamma_{\mu}, \sigma_{\mu \nu}\}$, where as usual $\sigma_{\mu \nu} \equiv i [\gamma_{\mu}, \gamma_{\nu}]/2$. Thus, we see that
\begin{equation}
\label{eq:normal1}
\langle : q_{a \alpha} (x) \bar{q}_{b \beta} (0) : \rangle = \delta_{ab} \left( a \delta_{\alpha \beta} + b_{\mu \nu} \sigma^{\mu \nu}_{\alpha \beta} \right),
\end{equation}
where $a$, $b_{\mu \nu}$ are determined below. In the vacuum, only the first term appears, by parity and time reversal invariance. However, the external magnetic field breaks time reversal invariance and, thus, the tensor term in (\ref{eq:normal1}) is now allowed. Since the only tensor at our disposal is the external electromagnetic field $F_{\mu \nu}$, one sees that $b_{\mu \nu} \propto F_{\mu \nu}$. The quantities in \eqref{eq:normal1} can be obtained by suitable contractions of both sides with the appropriate Dirac matrices, which gives
\begin{equation}
\label{eq:normal12}
\langle : q_{a \alpha} (x) \bar{q}_{b \beta} (0) : \rangle = -\frac{\delta_{ab}}{12} \left(\langle : \bar{q} (0) q(x) : \rangle \delta_{\alpha \beta} + \frac{1}{2} \langle : \bar{q} (0) \sigma_{\mu \nu} q(x) : \rangle \sigma^{\mu \nu}_{\alpha \beta} \right).
\end{equation}
Since we are performing a short distance expansion, we can Taylor expand the quark field $q_a(x)$ for small $x$. In the fixed point gauge for the color gauge potential, $x^{\mu} A_{\mu} = 0$, we have $x^{\mu} D_{\mu} = x^{\mu} \partial_{\mu}$. So, the Taylor expansion takes the form
\begin{equation}
\label{eq:expansion}
q_a (x) = q_a(0) + x^{\mu} D_{\mu} q_a |_{x=0} + \frac{1}{2} x^{\mu} x^{\nu} D_{\mu} D_{\nu} q_a |_{x=0} + \cdots
\end{equation}
Since we are interested only in dimension three operators, we can truncate the Taylor expansion to the zeroth order term since keeping higher order terms in the expansion corresponds to considering condensates of higher dimensions. With this expansion, we can write the non-perturbative part of the quark propagator (up to condensates of dimension 3) as
\begin{equation}
\label{eq:quarkOPE1}
\langle : q_{a \alpha} (x) \bar{q}_{b \beta} (0) : \rangle = -\frac{\delta_{ab}}{12} \langle : \bar{q} q : \rangle \delta_{\alpha \beta} - \frac{\delta_{ab}}{24} \langle : \bar{q} \sigma_{\mu \nu} q : \rangle \sigma^{\mu \nu}_{\alpha \beta}.
\end{equation}
In our case, the magnetic field is in the $z$ (3) spatial direction. In this situation, the only non-zero magnetic condensates are $\langle : \bar{q} \sigma_{1 2} q : \rangle$ and $\langle : \bar{q} \sigma_{2 1} q : \rangle = -\langle : \bar{q} \sigma_{1 2} q : \rangle$. Thus, in this case
\begin{equation}
\label{eq:quarkOPE2}
\langle : q_{a \alpha} (x) \bar{q}_{b \beta} (0) : \rangle = -\frac{\delta_{ab}}{12} \langle : \bar{q} q : \rangle \delta_{\alpha \beta} - \frac{\delta_{ab}}{12} \langle : \bar{q} \sigma_{1 2} q : \rangle \sigma^{1 2}_{\alpha \beta}.
\end{equation}

\section{Correlator in the weak field limit}

\begin{figure}
  \centering
  \includegraphics[width=1.1\textwidth]{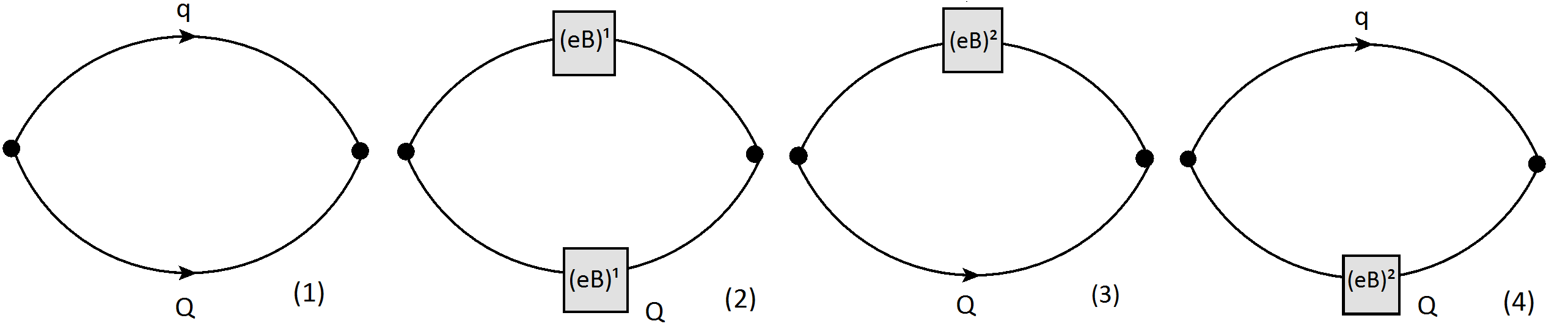}
  \caption{Perturbative diagrams up to order $(eB)^2$.}
  \label{fig:diag}
\end{figure}

The diagrams with non-vanishing traces are shown in Fig.\ \ref{fig:diag}. In this notation, the line with the square corresponds to the term $(eB/m^2)$ or $(eB/m^2)^2$ of the weak field propagator expansion in Eq.\ (\ref{eq:propweak}).

The diagram (1) in Fig.\ (\ref{fig:diag}) leads to
\begin{align}
\Pi_{(1)} &= 3\cdot 4 i \int\frac{d^4 k}{(2\pi)^4}\frac{k^2+k\cdot q -m_q m_Q}{(k^2-m_Q^2)((k+q)^2-m_q^2)} =  \nonumber\\
&= 3\cdot 4 i \int^1_0 dx \int \frac{d^4 k}{(2\pi)^4} \left[\frac{k^2}{(k^2 -\Delta)^2} + \frac{(-xq^2(1-x)-m_q m_Q)}{(k^2 -\Delta)^2} \right],
\end{align}
with $\Delta \equiv  -x(1-x)q^2 + x m_Q^2 + (1-x)m_q^2$, where $x$ is a Feynman parameter. After integrating over the momentum, we obtain
\begin{align}
\Pi_{(1)} =  \frac{3\cdot 4}{(4\pi)^2}\int_0^1 dx \ln \Delta \left(2\Delta - xq^2(1-x)-m_q m_Q \right).
\end{align}
This is the usual perturbative, $eB=0$, contribution to the correlator.

The diagram (2) in Fig.\ (\ref{fig:diag}) corresponds to
\begin{align}
\Pi_{(2)} 
= 3\cdot 4 i (cB)(CB) \int \frac{d^2 k_{\perp}}{(2\pi)^2} \int \frac{d^2 k_{\parallel}}{(2\pi)^2} \frac{k_{\parallel}^2 + k_{\parallel} \cdot q_{\parallel}+m_q m_Q}{((k+q)^2-m_q^2)^2(k^2-m_Q^2)},
\end{align}
with $k_{\parallel}=(k_0,k_3)$ and $k_{\perp}=(k_1,k_2)$. After the Feynman parametrization we arrive at
\begin{align}
\Pi_{(2)} = 3\cdot 4 i (cB)(CB) 6 \int_0^1 dx \,x(1-x)
\int \frac{d^2 k_{\perp}}{(2\pi)^2} \int \frac{d^2 k_{\parallel}}{(2\pi)^2}
\left[\frac{k_{\parallel}^2}{(k_{\parallel}^2-\Delta_{\parallel})^4} + \frac{(-xq_{\parallel}^2+x^2q_{\parallel}^2-m_q m_Q)}{(k_{\parallel}^2-\Delta_{\parallel})^4}\right],&
\end{align}
with $\Delta_{\parallel} = k_{\perp} -x(1-x)q_{\parallel}^2-x(1-x)q_{\perp}^2-x(-m_q^2+m_Q^2)+m_Q^2$. Then, we can evaluate the $k_{\parallel}$ integral. The same procedure applies to the $k_{\perp}$ integral. In the end, we obtain the following expression  

\begin{align}
\Pi_{(2)}& = 3\cdot 4 (cB)(CB) \frac{1}{(4\pi)^2} \int_0^1 dx x(1-x) \left[\frac{1}{\Delta_{\perp}} - \frac{(q_{\parallel}^2(x^2-x)-m_q m_Q)}{2\Delta_{\perp}^2}\right],
\end{align}
with $\Delta_{\perp}= (x^2-x)q^2+xm_q^2+(1-x)m_Q^2$.

The diagram (3) in Fig.\ (\ref{fig:diag}) is given by
\begin{align}
\Pi_{(3)}  = 3\cdot 4 i (CB)^2 \int\frac{d^4 k }{(2\pi)^4} \left[2\frac{(k_1^2+k_2^2)(m_q m_Q-k\cdot(k+q))}{((k+q)^2-m_q^2)(k^2-m_Q^2)^4} \right. + & \nonumber \\ \left. -2\frac{(k_1\cdot(k_1+q_1)+k_2\cdot(k_2+q_2))}{((k+q)^2-m_q^2)(k^2-m_Q^2)^3}\right]\,.
\end{align}

By the same procedure worked out for the diagram (2), we obtain
\begin{align}
\Pi_{(3)}=  \frac{3 \cdot 4}{(4\pi)^2} (CB)^2 \int^1_0 dx (1-x)^3 \left[\frac{1}{\Delta_{\perp}} + \frac{q^2(x-x^2)-q_{\perp}^2(x-4x^2)+m_q m_Q}{3\Delta_{\perp}^2} \right. + & \nonumber\\ +\left. \frac{2x^2q_{\perp}^2(q^2(x-x^2)+m_q m_Q)}{3\Delta_{\perp}^3} \right] -\frac{3\cdot 4}{(4\pi)^2} (CB)^2 \int^1_0 (1-x)^2 \left[\frac{1}{\Delta_{\perp}}+\frac{q_{\perp}^2(x^2-x)}{\Delta_{\perp}^2} \right],&
\end{align}
with $\Delta_{\perp} = (x^2-x)q^2 + xm_q^2 + (1-x)m_Q^2$. In order to obtain the result for the diagram (4) we can just do $C\leftrightarrow c$, $m\leftrightarrow m_Q$ and $(q+k)\leftrightarrow k$ in the previous result for $\Pi_{(3)}$.